\documentclass[11pt]{article}
\pdfoutput=1
\usepackage[textwidth = 430 pt, textheight = 630 pt]{geometry}
\usepackage{color}
\definecolor{MyDarkBlue}{rgb}{0.15,0.25,0.45}
\usepackage[linktocpage=true]{hyperref}
\hypersetup{
colorlinks=true,
citecolor=MyDarkBlue,
linkcolor=MyDarkBlue,
urlcolor=MyDarkBlue,
pdfauthor={Christian S\"amann and Richard J. Szabo},
pdftitle={Quantization of 2-Plectic Manifolds},
pdfsubject={hep-th}
}

\let\fn\footnote
\renewcommand{\footnote}[1]{\linespread{1.1}\fn{#1}\linespread{1.29}}

\usepackage{fancyhdr}
\usepackage[left]{lineno}

\makeatletter\renewcommand{\section}{\@startsection
{section}{1}{\z@}{-3.5ex plus -1ex minus
    -.2ex}{2.3ex plus .2ex}{\bf }}
\makeatletter\renewcommand{\subsection}{\@startsection{subsection}{2}{\z@}{-3.25ex
plus -1ex minus
   -.2ex}{1.5ex plus .2ex}{\it }}
\makeatletter\renewcommand{\subsubsection}{\@startsection{subsubsection}{3}{-2.45ex}{-3.25ex
plus -1ex minus -.2ex}{1.5ex plus .2ex}{\it }}

\makeatletter \@addtoreset{equation}{section}

\renewenvironment{thebibliography}[1]
     {\baselineskip=16pt plus 2pt minus 1pt
      \section*{\large\refname
        \@mkboth{\MakeUppercase\refname}{\MakeUppercase\refname}}%
     \list{\@biblabel{\@arabic\c@enumiv}}%
           {\settowidth\labelwidth{\@biblabel{#1}}%
            \leftmargin\labelwidth
            \advance\leftmargin\labelsep
            \@openbib@code
            \usecounter{enumiv}%
            \let\p@enumiv\@empty
            \renewcommand\theenumiv{\@arabic\c@enumiv}}%
      \sloppy
      \clubpenalty4000
      \@clubpenalty \clubpenalty
      \widowpenalty4000%
      \sfcode`\.\@m}

\usepackage{epsfig,rotating}
\usepackage{amsmath,amssymb}
\usepackage{amsfonts}
\usepackage{mathrsfs}
\usepackage{bbm}
\usepackage{bm}

\def\slasha#1{\setbox0=\hbox{$#1$}#1\hskip-\wd0\hbox to\wd0{\hss\sl/\/\hss}}

\def\periodb#1{\setbox0=\hbox{$#1$}#1\hskip-\wd0\hbox to\wd0{-}}

\newcommand{\delder}[1]{\frac{\delta}{\delta #1}}   		% partielle ableitung, 1 argument

\newcommand{\unit}{\mathbbm{1}}   			% identity map/matrix
\newcommand{\blbr}{\big(\hspace{-0.1cm}\big(}
\newcommand{\brbr}{\big)\hspace{-0.1cm}\big)}
\newcommand{\CA}{\mathcal{A}}    			% cal-letters

\newcommand{\xd}{\dot{x}}

\newcommand{\CC}{\mathcal{C}}

\newcommand{\CCH}{\mathscr{H}}

\newcommand{\CL}{\mathcal{L}}

\newcommand{\CN}{\mathcal{N}}
\newcommand{\CO}{\mathcal{O}}

\newcommand{\CP}{\mathcal{P}}

\newcommand{\CT}{\mathcal{T}}

\newcommand{\frg}{\mathfrak{g}}				% frak-letters

\newcommand{\FR}{\mathbbm{R}}     			% field of real numbers
\newcommand{\FC}{\mathbbm{C}}     			% field of complex numbers
\newcommand{\aosp}{\mathfrak{osp}}
\newcommand{\NN}{\mathbbm{N}}     			% set of natural numbers
\newcommand{\RZ}{\mathbbm{Z}}     			% ring of integers
\newcommand{\CPP}{{\mathbbm{C}P}}    			% complex projective plane
\newcommand{\ah}{\hat{a}}
\newcommand{\xh}{\hat{x}}

\newcommand{\fh}{\hat{f}}

\newcommand{\dd}{\mathrm{d}}     			% total differential
\newcommand{\dpar}{\partial}     			% partial differential
\newcommand{\di}{\mathrm{i}}     			% imaginary unit
\newcommand{\eps}{{\varepsilon}}			% antisymmetric tensors
\newcommand{\zb}{{\bar{z}}}

\newcommand{\eand}{{~~~\mbox{and}~~~}}     		% and etc. in equations

\newcommand{\efor}{{~~~\mbox{for}~~~}}

\newcommand{\der}[1]{\frac{\dpar}{\dpar #1}}   		% partielle ableitung, 1 argument
\newcommand{\dder}[1]{\frac{\dd}{\dd #1}}   		% partielle ableitung, 1 argument
\newcommand{\tr}{\,\mathrm{tr}\,}     			% trace
\newcommand{\au}{\mathfrak{u}}
\newcommand{\asu}{\mathfrak{su}}

\newcommand{\sU}{\mathsf{U}}     			% groups

\newcommand{\sSO}{\mathsf{SO}}

\newcommand{\sEnd}{\mathsf{End}\,}

\newcommand{\remark}[1]{}     				% remark
\def\tyng(#1){\hbox{\tiny$\yng(#1)$}}			% small Young diagram
\def\tyoung(#1){\hbox{\tiny$\young(#1)$}}			% small Young diagram
\newcommand{\Pair}{\mathrm{Pair}}

\begin{document}
\begin{flushright}
 HWM--11--13 \\ EMPG--11--17
\end{flushright}
\begin{center}
{\Large Quantization of $2$-Plectic Manifolds\footnote{Based on talk given by CS at the fourth annual meeting of the European Noncommutative Geometry Research Training Network in
Bucharest on 29 April 2011. To appear in the proceedings.}}\\[0.3cm] 
{\large Christian S\"amann and Richard J.\ Szabo}\\[0.3cm]
{\em Department of Mathematics and Maxwell Institute for Mathematical Sciences\\
Heriot-Watt University\\
Colin Maclaurin Building, Riccarton, Edinburgh EH14 4AS, U.K.}\\[5mm] {E-mail: {\ttfamily C.Saemann@hw.ac.uk , R.J.Szabo@hw.ac.uk}} 
\end{center}

\begin{quote}
We describe an extension of the axioms of quantization to the case of 2-plectic manifolds. We show how such quantum spaces can be obtained as stable classical solutions in a zero-dimensional 3-algebra reduced model obtained by dimensional reduction of the Bagger-Lambert-Gustavsson theory. We demonstrate that the groupoid approach to geometric quantization proposed by Hawkins and others can be useful for quantizing 2-plectic manifolds. We discuss our results in the context of recent developments in the quantum geometry of M-branes, and in the nonassociative deformation of spacetime induced by closed strings in the presence of a 2-plectic form.
\end{quote}

\section{Introduction}

A {\em symplectic manifold} $(M,\omega)$ is a manifold $M$ endowed with a globally defined non-degenerate closed 2-form $\omega$. That is, $\dd \omega=0$ and if the contraction $\iota_v \omega(x)=0$ for $v\in T_xM$, then $v=0$. The inverse to the matrix describing $\omega$ locally is a bivector field giving rise to a Poisson structure on $M$. Therefore, $M$ can serve as a phase space in Hamiltonian dynamics, and the quantization of such a phase space is a standard problem in quantum mechanics.

There is a natural generalization of symplectic manifolds known as {\em multisymplectic} or {\em $p$-plectic manifolds}. Here, one considers the pair $(M,\varpi)$ consisting of a manifold $M$ and  a globally defined non-degenerate closed $p+1$-form $\varpi$ on $M$. Again, non-degeneracy of $\varpi$ means that from $\iota_v \varpi(x)=0$ for $v\in T_xM$ it follows that $v=0$. The 1-plectic manifolds are the symplectic manifolds; we will be primarily interested in 2-plectic manifolds which are characterized by a 3-form $\varpi$. On a $p+1$-dimensional manifold, the $p$-plectic form is a volume form and can be inverted to give rise to a {\em Nambu-Poisson structure} on $M$. This is a $p+1$-ary bracket on $\CC^\infty(M)$ satisfying a generalized Leibniz rule 
\begin{equation}
 \{f_1 \,f_2,f_3,\dots ,f_{p+2}\}=f_1\,\{f_2,\dots
 ,f_{p+2}\}+\{f_1,f_3, \dots ,f_{p+2}\} \,f_2
\end{equation}
as well as the {\em fundamental identity}
\begin{eqnarray}
 \{f_1,\dots ,f_{p},\{g_1,\dots ,g_{p+1}\}\}&=& \{\{f_1,\dots ,f_{p},g_1\},g_2, \dots ,g_{p+1}\} \nonumber\\ && +\, \dots +\{g_1,\dots ,g_p,\{f_1,\dots ,f_{p},g_{p+1}\}\}
\end{eqnarray}
for $f_i,g_i\in\CC^\infty(M)$. Manifolds endowed with a Nambu-Poisson structure can be used as multiphase spaces in Nambu mechanics, and we are interested in the (higher) quantization of such multiphase spaces.

Our motivation for considering multisymplectic manifolds and their quantization stems from the description of a configuration of M2-branes ending on M5-branes that we review in the following. This configuration is the M-theory lift of the D-brane interpretation \cite{Diaconescu:1996rk,Tsimpis:1998zh} of Nahm's equations~\cite{Nahm:1979yw}. Consider the vacuum configuration of $k$ coincident D1-branes ending on a single D3-brane in type IIB superstring theory, with wrapped directions depicted schematically as~\cite{Diaconescu:1996rk,Tsimpis:1998zh}
\begin{equation}\label{diag:D1D3}
\begin{tabular}{rcccccccc}
& 0 & 1 & 2 & 3 & 4 & 5 & 6 \\
D1 & $\times$ & & & & & & $\times$ \\
D3 & $\times$ & $\times$ & $\times$ & $\times$ & & & 
\end{tabular}
\end{equation}
We work with cartesian coordinates $x^0,x^1,\ldots,x^6$ on $\FR^{1,6}$ and define $s=x^6$. The D3-brane is located at $s=0$. From the perspective of the D3-brane, the endpoint of the D1-branes looks precisely like a Dirac monopole. From the perspective of the D1-branes, the dynamics of this configuration are described by the Nahm equations
\begin{equation}\label{eq:Nahm}
 \dder{s}X^i=\tfrac{1}{2}\, \eps^{ijk}\, [X^j,X^k]~,
\end{equation}
where $X^i\in \au(k)$, $i,j,k=1,2,3$ describe the transverse fluctuations of the $k$ D1-branes parallel to the worldvolume directions of the D3-brane. This system is manifestly $\sSO(3)$-invariant under rotations in the spatial worldvolume directions of the D3-brane.

The simplest solution to the Nahm equations~\eqref{eq:Nahm} is found from a product ansatz $X^i=r(s)\, G^i$, which leads to
\begin{equation}\label{eq:fuzzyfunnel}
 r(s)= \frac{1}{s} \quad \eand \quad G^i=\eps^{ijk}\, [G^j,G^k]~.
\end{equation}
For technical reasons, the representation of $\asu(2)$ formed by the $G^i$ has to be irreducible. The $G^i$ can thus be considered as coordinates on a fuzzy sphere. This solution therefore suggests that each point in the worldvolume of the D1-branes polarizes into a fuzzy sphere $S_F^2$, gaining two spatial dimensions. Moreover, the radius $r(s)$ of these spheres diverges towards the position of the D3-brane at $s=0$. The solution \eqref{eq:fuzzyfunnel} is known as a {\em fuzzy funnel} \cite{Myers:1999ps} and it describes the smooth transition between $k$ D1-branes and a D3-brane with partially noncommutative worldvolume. The radial dependence $r=\frac{1}{s}$ matches the Higgs field $\Phi= \frac{1}{r}$ in the effective D3-brane description, where $\Phi$ is to be identified with $s$.

We can lift the configuration \eqref{diag:D1D3} to M-theory via T-duality along the $x^5$-direction and interpreting $x^4$ as the M-theory direction. This leads to the wrapped directions
\begin{equation}\label{diag:M2M5}
\begin{tabular}{rccccccc}
 & 0 & 1 & 2 & 3 & \phantom{(}4\phantom{)} & 5 & 6 \\
M2 & $\times$ & & & & & $\times$ & $\times$ \\
M5 & $\times$ & $\times$ & $\times$ & $\times$ & $\times$ & $\times$ 
\end{tabular}
\end{equation}
From the perspective of the M5-brane, the boundary of the coincident M2-branes is described by a {\em self-dual string} \cite{Howe:1997ue}. Basu and Harvey \cite{Basu:2004ed} suggested an equation for the description of the configuration \eqref{diag:M2M5} from the perspective of the M2-brane; it is given by
\begin{equation}\label{eq:BasuHarvey}
 \dder{s}X^\mu=\tfrac{1}{3!}\, \eps^{\mu\nu\kappa\lambda}\, [X^\nu,X^\kappa,X^\lambda]~, \qquad \mu,\nu,\kappa,\lambda=1,2,3,4~.
\end{equation}
This equation is a natural extension of the $\sSO(3)$-symmetric Nahm equation \eqref{eq:Nahm} describing the $\sSO(3)$-symmetric configuration \eqref{diag:D1D3} to the $\sSO(4)$-symmetric situation \eqref{diag:M2M5}. We assume that the triple bracket appearing in \eqref{eq:BasuHarvey} is a trilinear and totally antisymmetric map on some vector space. We call a vector space endowed with such a 3-bracket a 3-algebra\footnote{Usually, it is assumed that this 3-algebra is in fact a 3-Lie algebra \cite{Filippov:1985aa} whereby the 3-bracket obeys additional axioms. Here, we leave the definition intentionally open.}.

If we choose again a product ansatz $X^\mu=r(s)\, G^\mu$, then we find
\begin{equation}\label{eq:fuzzyfunnelM}
 r(s)= \frac{1}{\sqrt{2s}} \quad \eand \quad G^\mu=\eps^{\mu\nu\kappa\lambda}\, [G^\mu,G^\nu,G^\kappa]~.
\end{equation}
This solution matches the profile from a supergravity analysis. We would like to interpret this solution again as a fuzzy funnel; each point in the worldvolume of the M2-branes should polarize into a fuzzy three-sphere $S_F^3$, gaining three spatial dimensions. In this case a 3-form structure appears, and we require a clear interpretation of the quantization of the 2-plectic manifold $S^3$.

2-plectic manifolds further appear in M-theory from the perspective of the M5-brane; by turning on a constant 3-form $C$-field background, the self-dual strings propagate in a quantization of (two copies of) the 2-plectic space $\FR^3$ described by the 3-bracket relation $[x^\mu,x^\nu,x^\lambda]=\theta^{\mu\nu\lambda}$ with $\mu,\nu,\lambda=0,1,2$ and $\mu,\nu,\lambda=3,4,5$~\cite{Chu:2009iv}, where $\theta^{\mu\nu\lambda}$ is related to the 2-plectic form.

Further recent motivation stems from \emph{closed} string
theory. In~\cite{Baez:2008bu} it is shown that the phase space of the
bosonic string can be interpreted as a 2-plectic
manifold. In~\cite{Lust:2010iy} it is shown that for three-dimensional
backgrounds with 3-form flux, the coordinates of closed strings obey
the noncommutative relations $[x^i,x^j]=\theta^{ijk}\,\partial_k$,
where $\theta^{ijk}$ is related to the flux. In this case the Jacobi
identity is not satisfied and leads to a nonassociative 3-bracket
structure $[x^i,x^j,x^k]=\theta^{ijk}$~\cite{Blumenhagen:2010hj}. This
structure appears to be related to the quantization of a higher
Poisson structure on e.g. $\FR^3$ with multivector field
$\pi=\frac1{3!}\, \theta^{ijk}\,\partial_i\wedge \partial_j\wedge \partial_k$.

In the following, we review the results of \cite{DeBellis:2010pf} and \cite{DeBellis:2010sy}; in section 2 we propose a generalization of the quantization axioms to $p$-plectic manifolds, while in section 3 we illustrate how quantized 2-plectic manifolds arise dynamically as vacua of 3-algebra reduced models. In section 4 we present an extension of the groupoid approach to quantization suggested by Hawkins~\cite{Hawkins:0612363} to 2-plectic manifolds involving loop spaces, cf.\ also \cite{Saemann:2011zq}. In section 5 we close with some concluding remarks.

\section{Quantization axioms}

In geometric quantization, one starts from a symplectic manifold $M$ with a symplectic 2-form $\omega$ representing the first Chern class in $H^2(M,\RZ)$ of a principal $\sU(1)$-bundle. The associated vector bundle is the pre-quantum line bundle, and a Hilbert space is constructed from its global sections. 

Integral 2-plectic forms define the Dixmier-Douady class of an abelian gerbe, and one would expect this gerbe to take the role of the $\sU(1)$-bundle in a geometric quantization of 2-plectic manifolds. This, together with the expected appearance of nonassociativity, suggests that functions should no longer be mapped to endomorphisms on some Hilbert space. However, in the currently most promising candidate for an effective description of multiple M2-branes, the ABJM model~\cite{Aharony:2008ug}, all higher-bracket structures are replaced by matrix products. We therefore first examine how far a ``naive'' approach to the quantization of 2-plectic manifolds involving ordinary Hilbert spaces can take us.

\subsection{Quantization axioms for symplectic manifolds}

To find a suitable set of quantization axioms for 2-plectic manifolds, let us first recall the symplectic case. At classical level, states in a physical system are given by points on a Poisson manifold $M$. The observables in this system are given by smooth functions on $M$. At quantum level, states are rays in a complex Hilbert space $\CCH$ and observables are linear operators on $\CCH$. A {\em quantization of a Poisson manifold} $M$ therefore consists of a Hilbert space $\CCH$ together with a map $\widehat{-}:\CC^\infty(M)\rightarrow \sEnd(\CCH)$ satisfying certain conditions.

A {\em full quantization} is a quantization prescription which satisfies a comprehensive list of axioms:
\begin{enumerate}
\setlength{\itemsep}{-1mm}
 \item[Q1.] The quantization map $f\mapsto \hat{f}$ is linear over $\FC$ and maps real functions $f$ to hermitian operators $\hat{f}$.
 \item[Q2.] The constant function $f=1$ is mapped to the identity operator on $\CCH$: $\widehat{1}=\unit_\CCH$.
 \item[Q3.] The correspondence principle is satisfied: $\{f_1,f_2\}=g\ \Longrightarrow\ [\hat{f}_1,\hat{f}_2]=-\di \, \hbar\, \hat{g}$.
 \item[Q4.] The quantized coordinate functions act irreducibly on $\CCH$.
\end{enumerate}

While a full quantization of the two-torus $T^2$ can be constructed, this is unfortunately not true for such common spaces as the cotangent bundle $T^*\FR^n$ or the two-sphere $S^2$. It is therefore necessary to relax the axioms of a full quantization. Three ways have proved to be successful. First, one can drop the irreducibility condition Q4. Second, one defines the quantization map only on a subset of $\CC^\infty(M)$. Third, one demands that the correspondence principle Q3 only applies to first order in the parameter $\hbar$. The first two approaches yield {\em prequantization}, which is the starting point e.g.\ of {\em geometric quantization}. The third approach leads to {\em deformation quantization}.

In the following, we focus on {\em Berezin quantization}, which is better known as {\em fuzzy geometry} in the physics community and is a hybrid of geometric and deformation quantization. The reason is simply that in the D-brane configuration \eqref{diag:D1D3}, the Berezin quantized two-sphere appears, and our main motivation is to lift this picture to M-theory.

\subsection{Example: Berezin quantization of $S^2$}

The construction of the Hilbert space for the {\em Berezin quantized} or {\em fuzzy sphere} \cite{Berezin:1974du,Madore:1991bw} follows that for geometric quantization. That is, we start from the ample line bundle $\CO(k)$, $k\in \NN$ over $\CPP^1$. The Hilbert space $\CCH=\CCH_k$ is identified with the global holomorphic\footnote{This means that we work with K\"ahler polarization.} sections of $\CO(k)$. Recall that elements of $H^0(\CPP^1,\CO(k))$ are given by homogeneous polynomials of degree $k$ in the homogeneous coordinates $z_\alpha$, $\alpha=1,2$ of $\CPP^1$. This space is isomorphic to the $k$-particle Hilbert subspace in the Fock space of two harmonic oscillators described by creation and annihilation operators satisfying $[\hat{a}_\alpha,\hat{a}_\beta^\dagger]=\delta_{\alpha\beta}$, $\hat{a}_\alpha|0\rangle=0$. Altogether, we have
\begin{equation}
 \CCH_k\cong {\rm span} (z_{\alpha_1}\cdots z_{\alpha_k})\cong {\rm span}\big( \ah^\dagger_{\alpha_1}\cdots \ah^\dagger_{\alpha_k}|0\rangle \big)~.
\end{equation}
For any $z\in\CPP^1$, we construct the Rawnsley coherent states and the corresponding coherent state projector
\begin{equation}
 |z\rangle=\frac{1}{k!}\, \big(\zb_\alpha\, \ah_\alpha^\dagger \big)^k|0\rangle \quad \eand \quad \CP:=\frac{|z\rangle\langle z|}{\langle z | z \rangle}~.
\end{equation}
The projector $\CP$ is simultaneously an endomorphism on $\CCH_k$ and a function on $\CPP^1$. It therefore provides a bridge between the classical and the quantum world. We define the {\em lower Berezin symbol} $\sigma:\sEnd(\CCH_k)\rightarrow \CC^\infty(\CPP^1)$ by
\begin{equation}
 \sigma(\,\hat{f}\, ):=\tr_{\CCH_k}\big(\, \CP \, \hat{f}\, \big)~.
\end{equation}
One easily verifies that $\sigma$ is injective. On the {\em set of quantizable functions} $\Sigma:=\sigma(\sEnd(\CCH_k))$, we thus have an inverse operation, which yields the quantization map
\begin{eqnarray}
&& f(z)=f^{\alpha_1\dots\alpha_k\beta_1\dots \beta_k}\, \frac{z_{\alpha_1}\cdots z_{\alpha_k}\, \zb_{\beta_1}\cdots \zb_{\beta_k}}{|z|^{2k}} \nonumber\\ && \qquad \qquad \ \longmapsto\ \hat{f}=f^{\alpha_1\dots\alpha_k\beta_1\dots\beta_k}\, \frac{1}{k!}\, \ah^\dagger_{\alpha_1}\cdots \ah^\dagger_{\alpha_k}|0\rangle\langle 0|\ah_{\beta_1}\cdots \ah_{\beta_k}~.
\end{eqnarray}
This quantization map indeed satisfies our quantization axioms Q1, Q2, Q4, and Q3 to linear order in $\hbar=\frac{2}{k}$. For more details on Berezin quantization, see \cite{IuliuLazaroiu:2008pk} and references therein.

\subsection{Quantization axioms for $p$-plectic manifolds}

The problem of quantizing 2-plectic manifolds is notoriously difficult. Most attempts in the past focused on extending geometric quantization, and in \cite{Dito:1996xr} a consistent approach, the {\em Zariski quantization}, was found. The resulting quantization prescription seems however unsatisfactory from a physics perspective. 

Here, we try to extend Berezin quantization. As the corresponding quantization axioms are weaker than those of geometric quantization, we expect this to be simpler. As mentioned above, we try to push a naive approach, which still encodes observables as linear operators in $\sEnd(\CCH)$ on a complex Hilbert space $\CCH$. Given an ordinary quantization of a $p$-plectic manifold $M$ (arising, say, from an additional Poisson structure on $M$), we say that this is a quantization of $M$ as a $p$-plectic manifold if the following axioms are satisfied:
\begin{enumerate}
\setlength{\itemsep}{-1mm}
 \item[Q1.] The quantization map $f\mapsto \hat{f}$ is invertible, linear over $\FC$ and maps real functions $f$ to hermitian operators $\hat{f}$.
 \item[Q2.] The constant function $f=1$ is mapped to the identity operator on $\CCH$: $\widehat{1}=\unit_\CCH$.
 \item[Q3.] The correspondence principle is satisfied to first order in $\hbar$:
\begin{equation}
  \lim_{\hbar\rightarrow 0}\,\Big\| \frac{\di}{\hbar}\,\sigma\big([\fh_1,\ldots,\fh_{p+1}]\big)-\{f_1,\ldots,f_{p+1}\}\Big\|_{L^2}=0~,
\end{equation}
where $\sigma:\sEnd(\CCH)\rightarrow \CC^\infty(M)$ is the inverse of the quantization map.
\end{enumerate}
If $M$ is a symplectic manifold, then these axioms hold for Berezin quantization.

\subsection{Example: Quantization of $\FR^3$}

The simplest example of a 2-plectic manifold is $\FR^3$ endowed with the 2-plectic form $\varpi=\eps_{ijk}\, \dd x^i\wedge \dd x^j\wedge \dd x^k$, $i,j,k=1,2,3$. We can ``invert'' this 3-form to the Nambu-Poisson bracket
\begin{equation}
 \{f,g,h\}=\eps^{ijk}\, \left(\der{x^i}f\right)\, \left(\der{x^j}g\right)\, \left(\der{x^k}h\right)~.
\end{equation}
To quantize $(\FR^3,\varpi)$ as a 2-plectic manifold, we need to find a Hilbert space $\CCH$ and a quantization map which fulfills the Nambu-Heisenberg algebra relation
\begin{equation}\label{eq:commR3L}
{}[\hat{x}^1,\hat{x}^2,\hat{x}^3]=-\di\,\hbar\, \unit_\CCH~. 
\end{equation}
One such quantization is given by the space $\FR^3_\lambda$, which was first constructed in \cite{Hammou:2001cc}. We start from a fuzzy sphere with Hilbert space $\CCH_k=H^0(\CPP^1,\CO(k))$. On $\sEnd(\CCH_k)$, we define a 3-bracket through the totally antisymmetrized operator product 
\begin{equation}
{}[\xh^1,\xh^2,\xh^3]=\eps_{ijk}\, \xh^i\, \xh^j\, \xh^k=-\di\,\frac{6R^3_k}{k}\,\unit_{\CCH_k}~.         
\end{equation}
The radius of this fuzzy sphere is $R_k=\sqrt{1+\frac{2}{k}}~ \sqrt[3]{\frac{\hbar\, k}{6}}$. The space $\FR^3_\lambda$ is now obtained by a ``discrete foliation'' of $\FR^3$ by fuzzy spheres; the total Hilbert space $\CCH$ is given by the direct sum of the Hilbert spaces $\CCH_k$ and the total quantization of a function $f$ is given by the block-diagonal operator whose $k$-th block is the quantization of $f$ on the fuzzy sphere with radius $R_k$. One easily checks that this quantization satisfies our generalized quantization axioms \cite{DeBellis:2010pf}. The quantum commutation relations \eqref{eq:commR3L} are similar to those derived in \cite{Blumenhagen:2010hj} and in~\cite{Chu:2009iv}.

\section{Quantum 2-plectic manifolds from 3-algebra reduced models}

In this section, we review how various quantized symplectic manifolds form vacuum solutions of the IKKT matrix model \cite{Ishibashi:1996xs} with background fields. We also show how quantized 2-plectic manifolds analogously form solutions in a zero-dimensional 3-algebra reduced model~\cite{DeBellis:2010sy}.

\subsection{Classical solutions in the IKKT matrix model}

The (twisted) action of the IKKT model including masses $\mu_I$ and a 3-form background field $C$ similar to the BMN matrix model \cite{Berenstein:2002jq} is given by 
\begin{equation}\label{eq:actionIKKT}
 S=\tr_{N}\Big(\big([X^I,X^J]-\theta^{IJ}\, \unit_N \big)^2+\mu_I \, (X^I)^2+C_{IJK}\, X^I\, [X^J,X^K]+\mbox{fermions}\Big)~,
\end{equation}
where $X^I\in\au(N)$, $I=0,1,\ldots,9$. For $\mu_I=C_{IJK}=0$ and $N\rightarrow \infty$, solutions to the classical equations of motion of \eqref{eq:actionIKKT} are given by the Moyal spaces $(\FR^{1,9},\theta^{IJ})$ with $[X^I,X^J]=\di\, \theta^{IJ}\, \unit$. In particular, one solution is given by the Moyal plane $\FR^2_\theta$. These solutions are stable and form global minima of the action \eqref{eq:actionIKKT}. Moreover, they are BPS solutions preserving half of the 32 supersymmetries of the IKKT model.

Turning on the 3-form background field $C_{123}=1$ and putting $\theta^{IJ}=0$, we obtain a fuzzy sphere solution $S_F^2$ with scalar fields obeying $[X^i,X^j]= \eps^{ijk}\, X_k$, $i,j,k=1,2,3$ and $X^I=0$ for $I\neq 1,2,3$. The solution is again both stable and BPS.

By turning on additional mass terms $C_{123}=1$, $\mu_1=\mu_2=\mu$, we find the stable (non-BPS) solution
\begin{equation}
 X^0=X^4=\ldots=X^9=0~, \qquad [X^1,X^2]= \theta^{12}\, \unit~, \qquad [X^3,X^i]= \theta^{ij}\, X^j
\end{equation}
for $i,j=1,2$. The Lie algebra formed by $X^1,X^2,X^3,\unit$ is called the {\em Nappi-Witten algebra}; it is identical to the linear Poisson structure on a four-dimensional Hpp-wave\footnote{This space is a four-dimensional Cahen-Wallach symmetric spacetime, see also \cite{Blau:2002mw,Halliday:2006qc,Rivelles:2002ez} for further details.}. We can therefore interpret this solution as a quantum Hpp-wave.

\subsection{Classical solutions in a 3-algebra reduced model}

The IKKT matrix model is obtained by dimensional reduction of a maximally supersymmetric Yang-Mills theory, e.g.\ the effective field theory of D2-branes, to zero dimensions. It is therefore natural to consider a corresponding reduction of the BLG model~\cite{Bagger:2007jr,Gustavsson:2007vu}. The latter is a Chern-Simons matter theory, which might provide an effective description of two M2-branes. Its reduction should take over the role of the IKKT model.

The field content of the BLG model consists of eight scalar fields $X^I$ as well as their fermionic superpartners $\Psi$ taking values in a metric 3-Lie algebra $\CA$. Additionally, one has a topological gauge potential $A_\mu$ taking values in the Lie algebra of inner derivations $\frg_\CA$ of $\CA$. Allowing again for mass terms and a 4-form field background, the dimensionally reduced action reads as 
\begin{equation}\label{eq:action3AM}
\begin{aligned}
 S=&-\tfrac{1}{2}\,\big(A_\mu X^I,A^\mu X^I\big)+\tfrac{\di}{2}\, \big(\bar{\Psi}, \Gamma^\mu\, A_\mu \Psi\big)
 -\tfrac{1}{2}\, \sum_{I=1}^8\, \mu^2_{1,I}\, \big(X^I,X^I\big)\\&+\tfrac{\di}{2}\, \mu_2\, \big(\bar{\Psi},\Gamma_{3456}\Psi \big)
 +C_{IJKL}\,\big([X^I,X^J,X^K],X^L\big)\\
&+\tfrac{\di}{4}\,\big(\bar{\Psi}, \Gamma_{IJ}[X^I,X^J,\Psi]\big)-\tfrac{1}{12}\,\big([X^I,X^J,X^K],[X^I,X^J,X^K]\big)\\
&+\tfrac{1}{6}\, \epsilon^{\mu \nu \lambda}\, \blbr A_\mu,[A_\nu,A_\lambda]\brbr+\frac{1}{4\gamma^2}\, \blbr[A_\mu,A_\nu],[A^\mu,A^\nu]\brbr~,
\end{aligned}
\end{equation}
where $(\cdot,\cdot)$ and $\blbr\cdot,\cdot\brbr$ denote the inner products on $\CA$ and $\frg_\CA$, respectively.

In the case $\mu_{1,I}=C_{IJKL}=0$, matter fields in a 3-algebra satisfying $[X^i,X^j,X^k]= \eps^{ijk}\, \unit$, $i,j,k=1,2,3$ form a stable BPS solution of the classical equations of motion of \eqref{eq:action3AM}. As we saw before, this solution can be interpreted as the noncommutative space $\FR^3_\lambda$.

If we turn on the background field $C_{1234}=1$, we find that stable BPS solutions are given by $[X^\mu,X^\nu,X^\kappa]=\eps^{\mu\nu\kappa\lambda}\, X^\lambda$, $\mu,\nu,\kappa,\lambda=1,2,3,4$. This solution corresponds to a fuzzy $S^3$, cf.\ \cite{DeBellis:2010pf} and references therein.

Giving additional mass terms to the scalar fields $X^1$ and $X^2$ by setting $\mu_{1,I}=\mu_2=\mu$, we find a stable solution if the matter fields satisfy a 3-algebra generalization  of the Nappi-Witten algebra given by
\begin{equation}
 [X^1,X^2,X^3]=\theta^{123}\, \unit \quad \eand \quad [X^4,X^i,X^j]=\theta^{ijk}\, X^k~.
\end{equation}
We interpret this solution as a noncommutative five-dimensional Hpp-wave background.

We have thus seen that a number of classes of solutions known from the IKKT matrix model is also found in a zero-dimensional 3-algebra reduced model. Recall that the BLG model can be reduced to three-dimensional maximally supersymmetric Yang-Mills theory by taking a strong coupling limit after Higgsing the model as proposed in \cite{Mukhi:2008ux}. Correspondingly, the IKKT matrix model \eqref{eq:actionIKKT} can be obtained from the 3-algebra model \eqref{eq:action3AM}. Therefore, the solutions of the IKKT model, i.e.\ the Moyal plane $\FR^2_\theta$, the fuzzy sphere $S^2_F$ and the noncommutative four-dimensional Hpp-wave, can be interpreted as strong coupling limits of the solutions of the 3-algebra model, i.e.\ $\FR^3_\lambda$, the fuzzy sphere $S^3_F$ and the noncommutative five-dimensional Hpp-wave.

There is a further connection to the IKKT matrix model. It was conjectured in~\cite{Azuma:2001re} that this model can be obtained from cubic matrix models whose gauge algebras are tensor products with the superalgebra $\aosp(1|32)$ \cite{Smolin:2000kc}. Our 3-algebra reduced model \eqref{eq:action3AM} shows also signs of an $\aosp(1|32)$-invariance. Some more details of the relation between the 3-algebra model and the $\aosp(1|32)$-matrix model are worked out in \cite{DeBellis:2010sy}.

Expanding the IKKT model around a solution corresponding to a quantized symplectic manifold yields a noncommutative supersymmetric gauge theory on this manifold. Expanding the 3-algebra model \eqref{eq:action3AM} around a solution, one obtains the BLG theory on the corresponding quantized 2-plectic manifold with additional higher derivative and other terms \cite{DeBellis:2010sy}. 

\section{Quantization via groupoids}

A {\em groupoid} is a small category in which every morphism is an isomorphism. More concretely, a groupoid is given by a set of objects and a set of composable, invertible arrows between these objects. Groupoids have been widely used in the context of noncommutative geometry and $C^*$-algebras; here we follow the approach of Hawkins \cite{Hawkins:0612363}. The motivation for using groupoids in the quantization of arbitrary Poisson manifolds stems from the observation that the quantization of the dual of a Lie algebra $\frg$ is straightforward and yields the twisted convolution algebra of the Lie group integrating $\frg$. As every Poisson manifold $M$ is naturally a Lie algebroid, it is very tempting to assume that the quantization of $M$ is given by a twisted convolution algebra of the Lie groupoid integrating $M$; this approach works in many examples. For us, an advantage of the groupoid approach to quantization is that it avoids the introduction of Hilbert spaces and cuts directly to the abstract $C^*$-algebra. It might therefore be able to lead to quantized 2-plectic manifolds involving nonassociative operator algebras.

Given a Poisson manifold $M$, the procedure proposed in \cite{Hawkins:0612363} consists of the following steps:
\begin{enumerate}
\setlength{\itemsep}{-1mm}
 \item Find an integrating symplectic groupoid $s,t:\Sigma\rightrightarrows M$ for the Lie algebroid $M$.
 \item Construct a prequantization of $\Sigma$ as in geometric quantization.
 \item Endow $\Sigma$ with a groupoid polarization.
 \item The quantization of $M$ is given by the polarized convolution algebra, twisted by the prequantum line bundle.
\end{enumerate}

\subsection{Groupoid quantization of $\FR^2$}

Let us sketch the simplest example for a groupoid quantization as presented in \cite{Hawkins:0612363}, the Poisson manifold $V=\FR^2$ with constant Poisson structure $\theta^{ij}$, $i,j=1,2$.

As an integrating groupoid, we take the pair groupoid $\Sigma=\Pair(V)\cong V\times V^*$ described by coordinates $(x^i,p_i)$ on $\Sigma$. The symplectic structure on $\Sigma$ is given by $\omega=\dd x^i\wedge \dd p_i$. The object inclusion map is trivially given by $\unit_x:(x^i)\mapsto(x^i,x^i)$, where we used the isomorphism between $V$ and its dual. We choose the source and target maps to be the Bopp shifts
\begin{equation}
 s(x^i,p_i)=(x^i+\tfrac{1}{2}\, \theta^{ij}\, p_j) \quad \eand \quad
 t(x^i,p_i)=(x^i-\tfrac{1}{2}\, \theta^{ij}\, p_j)~,
\end{equation}
and one easily verifies that $t$ is indeed a Poisson map, i.e. $\{t^*f,t^*g\}_{\omega^{-1}}=t^*\{f,g\}_\theta$, the first condition for $\Sigma$ to be a groupoid integrating $\FR^2$. Consider now the concatenation of arrows given by
\begin{equation}
 x^i+\theta^{ij}\, (p_j+p'_j)\ \longrightarrow \ x^i+\theta^{ij}\, (p_j-p'_j)\ \longrightarrow\ x^i-\theta^{ij}\, (p_j+p'_j)~.
\end{equation}
We can thus identify the set of composable arrows (the 2-nerve of $\Sigma$) with $V\times V^*\times V^*$. On this set, there are projections onto the first and second arrows given by
\begin{equation}
 \pi_1(x^i,p_i,p'_i\,)=(x^i+\tfrac{1}{2}\, \theta^{ij}\, p_j,p'_i\,)\quad \eand \quad
 \pi_2(x^i,p_i,p'_i\,)=(x^i-\tfrac{1}{2}\,\theta^{ij}\,p_j',p_i)~,
\end{equation}
as well as a multiplication of arrows
\begin{equation}
m(x^i,p_i,p'_i\, )=(x^i,p_i+p'_i\, )~. 
\end{equation}
Together these maps obey the consistency relations
\begin{equation}
\begin{aligned}
 t\big(\pi_1(x^i,p_i,p'_i)\big)=s\big(\pi_2(x^i,p_i,p'_i)\big)~,\hspace{3.7cm}&\\[4pt] s\big(m(x^i,p_i,p'_i)\big)=s\big(\pi_1(x^i,p_i,p'_i)\big) \quad \eand \quad t\big(m(x^i,p_i,p'_i)\big)=t\big(\pi_2(x^i,p_i,p'_i)\big)~.& 
\end{aligned}
\end{equation}
The second condition for $(\Sigma,\omega)$ to be an integrating symplectic groupoid for $V$ is that $\omega$ satisfies the multiplicativity property
\begin{equation}
 \dpar^*\omega:=\pi_1^*\omega-m^*\omega+\pi_2^*\omega=0~,
\end{equation}
which is indeed the case here.

The prequantization of $\Sigma$ is trivial, as $\Sigma\cong \FR^4$; the prequantum line bundle is the trivial line bundle $\Sigma\times \FC$ with connection of curvature $F=-2 \pi\, \di\, \omega$. The twist element $\sigma_0$ is found from the symplectic potential $\vartheta=-x^i\, \dd p_i$ on $\Sigma$, which also gives rise to the groupoid polarization $\CP$ of $\Sigma$ corresponding to the vector fields along the leaves of the fibration $\Sigma\rightarrow V^*$. The twist element is found from the equation
\begin{equation}
 \dpar^*\vartheta=\sigma_0^{-1}\, \dd\sigma_0=\dd(-\tfrac{1}{2}\, \theta^{ij}\, p_i\, p_j')~.
\end{equation}
We therefore have
\begin{equation}
\sigma_0=\exp\big(-\tfrac{1}{2}\, \theta^{ij}\, p_i\, p_j'\big) \ ,
\end{equation}
and the twisted polarized convolution algebra on $\Sigma/\CP\cong V^*$ is the algebra of functions on $\FR^2$ with the usual Moyal star-product.

\subsection{Groupoid quantization of $\FR^3$ using loop space}

To extend the above construction to the 2-plectic manifold $\FR^3$,
one is naturally led to a categorified approach involving
2-groupoids. Here, however, we follow an alternative approach
involving loop space. By using a transgression map \cite{0817647309}
(see also~\cite{Saemann:2011zq}), we reduce the 2-plectic structure on $\FR^3$ to a symplectic 2-form on the loop space $\CL \FR^3$ of $\FR^3$; here by $\CL \FR^3$ we mean the loop space of $\FR^3$ with reparametrizations factored out. 

Starting from the 2-plectic manifold $(\FR^3,\varpi)$ with $\varpi=\eps_{ijk}\, \dd x^i\wedge \dd x^j\wedge \dd x^k$, we construct the infinite-dimensional symplectic manifold $(\CL \FR^3,\CT\varpi)$ with symplectic form
\begin{equation}
 \CT\varpi:=\oint\, \dd \sigma\ \oint\, \dd \tau \ \varpi_{ijk}\, \xd^k(\tau)\, \delta(\tau-\sigma)\, \delta x^i(\sigma)\wedge \delta x^j(\tau)~.
\end{equation}
Note that $\CT\varpi$ is indeed non-degenerate; while we have
\begin{equation}
 \iota_X (\CT\varpi)=0 \quad \efor X=\oint\, \dd \sigma\ \xd^i(\sigma)\, \delder{x^i(\sigma)}~,
\end{equation}
this relation merely shows that $\CT\varpi$ is invariant under reparametrizations. To factor out reparametrization invariance, we can restrict ourselves to loops $x^i(\sigma)$, $\sigma\in [0,2\pi)$ with arc length parameterization $|\xd|=1$. We can then invert $\CT\varpi$ to find a Poisson structure. 

More generally, we can start from $\CL\FR^3$ with Poisson structure
\begin{equation}
 \{f,g\}:=\oint\, \dd \tau\ \oint\, \dd \sigma\ \delta(\tau-\sigma) \, \theta^{ijk}\, \xd_k(\sigma)\, \left(\delder{x^i(\tau)}f\right)\, \left(\delder{x^j(\sigma)}g\right)~,
\end{equation}
and follow the recipe of the previous subsection. That is, we choose the pair groupoid $\Sigma=\CL\FR^3\times \CL\FR^3$ with symplectic form $\omega=\oint\, \dd \tau\ \oint\, \dd \sigma\ \delta(\tau-\sigma)\, \delta x^i(\tau)\wedge \delta p_i(\sigma)$ as integrating groupoid. The source and target maps are
\begin{equation}
 \begin{aligned}
 s\big(x^i(\sigma)\,,\,p_i(\sigma)\big)&=x^i(\sigma)+\tfrac{1}{2}\, \theta^{ijk}\, p_j(\sigma)\, \xd_k(\sigma)~,\\[4pt] t\big(x^i(\sigma)\,,\,p_i(\sigma)\big)&=x^i(\sigma)-\tfrac{1}{2}\, \theta^{ijk}\, p_j(\sigma)\, \xd_k(\sigma)~,
 \end{aligned}
\end{equation}
where $x^i(\sigma)$ denotes a point in $\CL\FR^3$ given by the map
$x^i(\sigma): S^1\rightarrow \FR^3$, and so on. The 2-nerve of
$\Sigma$ is $\CL\FR^3\times \CL\FR^3\times \CL\FR^3$. One again checks that $\Sigma$ is an integrating symplectic groupoid by verifying that $t$ is a Poisson map and that $\omega$ satisfies $\dpar^*\omega=0$.

In this way, we arrive at a twisted polarized convolution algebra which is the algebra of functions on $\CL \FR^3$ with commutator
\begin{equation}
\big[x^i(\sigma)\,,\,x^j(\tau) \big]=\theta^{ijk}\, \delta(\sigma-\tau)\, \frac{\xd_k}{|\xd|}~.
\end{equation}
This result has been derived from a purely M-theory perspective in \cite{Bergshoeff:2000jn,Kawamoto:2000zt}. Moreover, when identifying $p_k$ with $\xd_k$, this relation is similar to that obtained in \cite{Lust:2010iy}; again, the 3-bracket is the failure of the Jacobi identity. The advantage of this approach is that the quantity $\theta^{ijk}\, \xd_k$ behaves like an ordinary Poisson structure, and this example reduces to $\FR^2$ by setting $\xd_k(\tau)=\delta_{k,3}\, \tau$.

\section{Conclusions} 

In this article we showed that a naive approach to the quantization of 2-plectic manifolds works surprisingly well. As a non-trivial test, we wrote down a 3-algebra reduced model, which contains the IKKT matrix model in a strong-coupling limit in the same way that the BLG model contains supersymmetric Yang-Mills theory in a strong coupling limit. Moreover, three classes of noncommutative spaces which appear as stable solutions of the IKKT matrix model can be obtained as the strong coupling limits of corresponding quantized 2-plectic manifolds.

A better motivated approach to the quantization of 2-plectic manifolds is, however, given by a generalization of the groupoid approach of Hawkins involving loop spaces. This approach reproduced relations found earlier in studies of M-theory and closed string theory.

Our results in the extension of groupoid quantization are only preliminary. Besides extending the discussion to other manifolds like $S^3$, it is also desirable to develop the groupoid quantization picture involving 2-groupoids. In this context, the interrelations and usefulness of various higher bracket structures (such as e.g.\ higher Poisson structures and Courant algebroids) should be clarified. Eventually, we hope to be able to rewrite the BLG model using an appropriate algebra of functions on $S^3$ as suggested by the M-brane configuration \eqref{diag:M2M5}.

\section*{Acknowledgments}

CS would like to thank the organizers of the EU-NCG Meeting for the invitation and the kind hospitality. The work of CS was supported by a Career Acceleration Fellowship from the UK Engineering and Physical Sciences Research Council. The work of RJS was supported in part by grant ST/G000514/1 ``String Theory
Scotland'' from the UK Science and Technology Facilities Council.

\end{document}